\documentclass[journal=jacsat,manuscript=article]{achemso}
\usepackage{graphicx}
\usepackage{amsmath,amssymb}
\usepackage{hyperref}
\usepackage{chemformula} 
\usepackage[T1]{fontenc} 

\author{Atanu Paul}
\affiliation[Bar-Ilan University]
{Department of Chemistry, Bar-Ilan University, Ramat Gan 5290002, Israel}

\author{Ilya Grinberg}
\affiliation[Bar-Ilan University]
{Department of Chemistry, Bar-Ilan University, Ramat Gan 5290002, Israel}

\email{ilya.grinberg@biu.ac.il}

\title[An \textsf{achemso} demo]  {An atomistic approach for modeling of polarizability and Raman scattering of water clusters and liquid water}

\abbreviations{IR,NMR,UV}

\begin{document}

\begin{tocentry}
\includegraphics[width = 8.65 cm,angle =0]{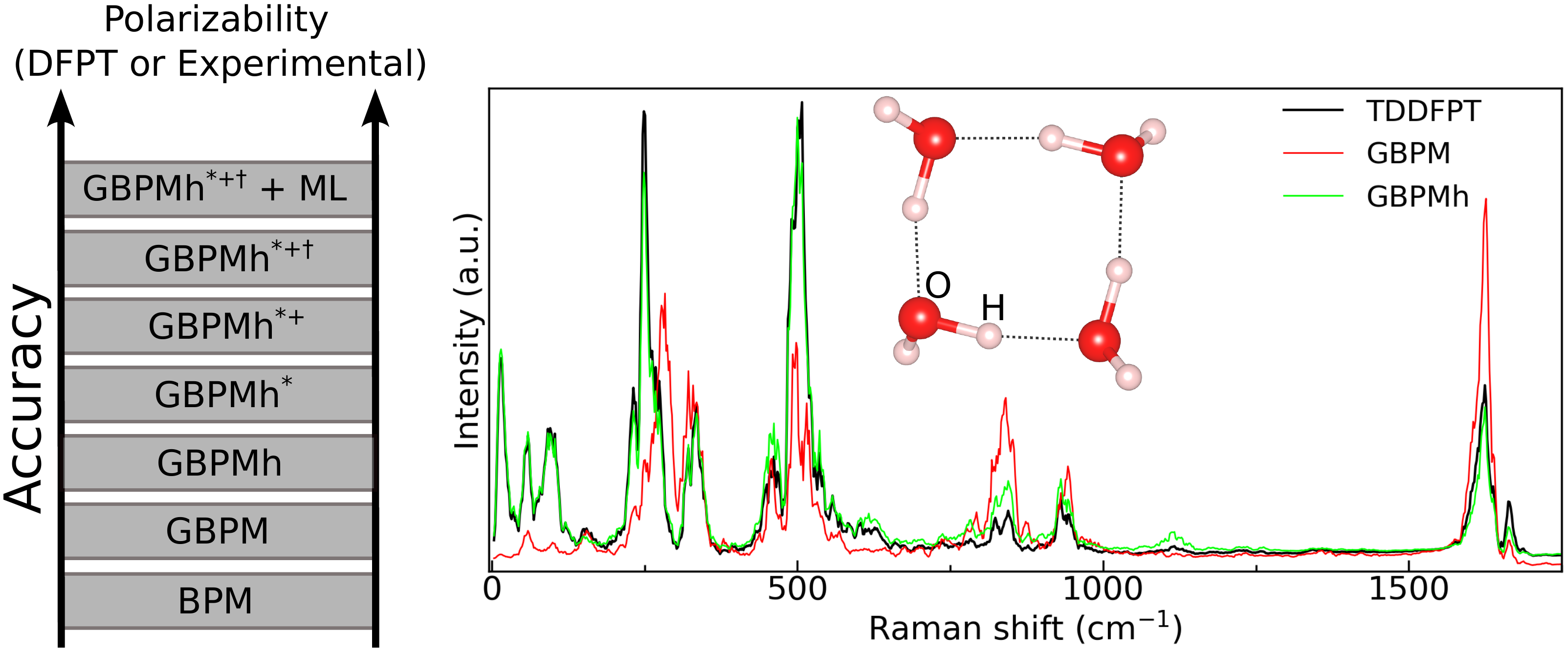}

\end{tocentry}

\begin{abstract}
In this work, we develop a framework for atomistic modeling of electronic polarizability to predict the Raman spectra of hydrogen-bonded clusters and liquids from  molecular dynamics (MD) simulations. The total polarizability of the system is assumed to arise from contributions of both the monomer unit and intermolecular interactions. The generalized bond-polarizability model (GBPM), inspired by the classic bond-polarizability model, effectively describes the electronic polarizability of a monomer. To account for the electronic polarizability arising from intermolecular interactions, we use a basis set of rapidly decaying functions of interatomic distances. We apply this model to calculate the electronic polarizability and Raman spectra of water clusters ((H$_{2}$O)$_{r}$, $r$ = 2, 3, 4, 5, 6) and liquid water. The computational results are compared with the results of quantum-mechanical  calculations for clusters and to experimental data for the liquid. It is demonstrated that this simple and physically motivated model, which relies on a small number of parameters, performs well for clusters at both low and high temperatures, capturing strong anharmonic effects. Moreover, its high transferability suggests its applicability to other water clusters. These results suggest that a hierarchical approach based on the Jacob’s ladder of  increasingly sophisticated and accurate atomistic polarizability models incorporating additional effects can be used for efficient modeling of Raman spectra from MD simulations of clusters, liquids and solids.
\end{abstract}

\section{Introduction}
Raman scattering based on inelastic light-matter interactions is an essential technique for probing the vibrational and structural 
properties of  gas-phase molecules, liquids and solids.~\cite{long,Raman_lipids,Ferrari2013} Theoretically, the positions of Raman peaks can be obtained by first-principles calculations at 0~K through vibrational mode (phonon) calculations; however, this is limited to the harmonic approximation and cannot model the temperature effects.~\cite{PhysRevLett.90.036401,UMARI20051255} The modeling of the evolution of Raman spectra with temperature requires the use of molecular dynamics (MD) simulations. Once the MD trajectory is known, the Fourier transform of the auto-correlation function of the fluctuations of the electronic polarizability ($\pmb{\alpha}$) along the MD trajectory can be used to obtain the Raman spectrum including both harmonic and anharmonic effects.~\cite{Berne,PhysRevLett.88.176401}   While the  electronic polarizability trajectory along the MD trajectory of the system can be calculated using density functional perturbation theory (DFPT),~\cite{PhysRevB.55.10355,DFPT_baroni} such calculations are very expensive and are limited to short trajectories and small system sizes. Therefore, atomistic modeling of electronic polarizability has attracted significant attention in recent work using both machine-learning-based and analytical model approaches.~\cite{car,Ceriotti,berger2024polarizability,acs.jpcc.4c00886,THOLE1981341,Thole_1,TholeL,polynomial}  In particular, the traditional analytical model approach has the advantages of being simple to interpret, computationally less expensive, and transferable in a straightforward fashion to larger systems.

The bond-polarizability model (BPM),~\cite{BPM1,1996273} is the simplest and oldest atomistic model of electronic polarizability  based only on the changes in the  length and orientation of covalent bonds. Despite its simplicity, this model has been applied successfully to various systems, particularly for the bond stretching modes in the high-wavenumber region of the spectrum.~\cite{PhysRevLett.104.085503,PhysRevB.63.094305,PhysRevB_241402,atanu_BPM} In our recent work, we have developed the  generalized bond-polarizability model (GBPM) that shows improved accuracy by taking into account the effect of the changes in the interatomic distances between atoms that are traditionally considered non-bonding (e.g. two H atoms in H$_{2}$O) but which in fact play an important role in both occupied and unoccupied frontier molecular orbitals that control the electronic polarizability.~\cite{GBPM}  This model shows essentially exact agreement with the DFPT results for the spectra of molecules, but unsurprisingly cannot reproduce the spectra of liquid water where intermolecular interactions play an important role. A similar incomplete accuracy was also found for the GBPM spectra of solid state systems such as the BaTiO$_{3}$ complex oxide.

Liquid water is known to be a highly complex system in which hydrogen bonds (H-bonds) play a key role in the dynamics.  H-bonds are much weaker than covalent bonds so that their vibrations are much more anharmonic and in liquid water H-bonds are constantly broken and reformed.  This means that a model for the electronic polarizability changes of liquid water must take a wide range of H-bond distances and configurations into account, which at first glance makes the task of atomistic modeling of the electronic polarizability contributions of H-bonds to be quite challenging. To date, this challenge was only addressed by Car and co-workers who developed a deep neural network (NN) model for polarizability and used it together with their deep potential for water to simulate the spectra of liquid water, obtaining good agreement with experimental spectrum.~\cite{car,car_water_model}  In their work, the deep NN model was parameterized using large-scale density functional theory (DFT) simulations of liquid water with Wannier function shifts under field providing the data to simulate the changes in the polarizability of individual water molecules in different local environments.  To obtain this dataset for parameterization, large and computationally expensive calculations are required. Additionally, the obtained deep NN model has a large number of parameters,
necessitating the use of heavy computational resources for simulations on large length and time scales.

Here, we seek to develop an analytical  atomistic  model for the calculations of electronic polarizability of water including H-bond effects using a small parameterization database. We use small water clusters ((H$_{2}$O)$_{r}$, $r$ = 2, 3, 4, 5, 6)  as our model systems because on the one hand, the small size of these clusters enables direct DFT MD simulations and time-dependent density-functional perturbation theory (TDDFPT) calculations of $\pmb{\alpha}$ along the MD trajectory, while on the other hand, these cluster contain H-bonds that are similar to the H-bonds in liquid water.  We show that a simple 54-parameter
analytical model can reproduce the TDDFPT Raman spectra of the clusters and obtains improved spectra of liquid water for MD simulations carried out using the AMOEBA model.~\cite{AMOEBA1,AMOEBA2,AMOEBA3}
The simplicity of the model allows easy interpretation and assignment of the spectra and its success in reproducing the complex spectra of water clusters suggests that the approach of systematically including effects of more complex interactions can be used to construct a Jacob's ladder of increasingly sophisticated and accurate analytical polarizability models.

\section{Theory and computational details}
In BPM, the total polarizability of the system is assumed to be composed of the polarizability  contributions of the individual bonds, i.e.,
\begin{equation}
\pmb{\alpha} = \sum_{k} \pmb{\alpha}^{k}
\label{eq1}
\end{equation}
where $k^{th}$ bond polarizability, $\pmb{\alpha}^{k}$, can be defined as:
\begin{equation}
 \pmb{\alpha}^{k} = \pmb{R}^{T}.\pmb{\alpha}^{k}_{local}.\pmb{R}
 \label{eq2}
\end{equation}
where $\pmb{\alpha}^{k}_{local}$ is the bond polarizability in the local Cartesian coordinate system and can be expressed in the pure diagonal form,
\begin{equation}
 \pmb{\alpha}^{k}_{local} = 
  \begin{bmatrix}
                 \alpha^{k}_{x^{'}} &         0            & 0                  \\
                         0          & \alpha^{k}_{y^{'}}   & 0                  \\
                         0          &         0            & \alpha^{k}_{z^{'}} \\
                \end{bmatrix}  
                \label{eq3}
\end{equation}
Here, $\pmb{R}$ is the rotation matrix of the coordinate transformation from the local bond coordinate system to the global Cartesian
coordinate system and $\pmb{R}^{T}$ is the transpose matrix of $\pmb{R}$. $\alpha^{k}_{x^{'}}$, $\alpha^{k}_{y^{'}}$ and $\alpha^{k}_{z^{'}}$ are the polarizability contributions due to the responses of the bond electron density  parallel (direction $\hat{x^{'}}$) and in the two perpendicular directions to the bond (directions $\hat{y^{'}}$ and $\hat{z^{'}}$), respectively. $\hat{x^{'}}$, $\hat{y^{'}}$ and $\hat{z^{'}}$ are the axes in the local bond coordinate system. We can also assume that the two perpendicular components have similar contributions to the bond polarizability due to the cylindrical nature of bonds. 

To calculate the total polarizability of the clusters or liquids, we can assume that the total polarizability of the systems is a sum of two contributions: intramolecular and intermolecular. To model the polarizability arising from intramolecular interactions, we employed the GBPM, where parallel ($\alpha^{k}_{x^{'}}$) and perpendicular ($\alpha^{k}_{y^{'}}$ or $\alpha^{k}_{z^{'}}$) bond polarizabilities are functions of the bond length ($\ell$) for all  bonding atom pairs and for certain specific non-bonding atom pairs. Then, the parallel and perpendicular bond polarizabilities for all considered atom pairs  can be expanded in the Taylor series around the equilibrium bond length ($\ell^{0}$),

\begin{equation}
 \alpha^{k}_{i^{'}} = \alpha^{0}_{i^{'}} + \alpha^{1}_{i^{'}} (\ell - \ell^{0}) + \alpha^{2}_{i^{'}} (\ell - \ell^{0})^{2} + ..
\label{eq4}
\end{equation}

where $\alpha^{k}_{i^{'}}$ is either parallel ($i^{'}$ = $x^{'}$) or perpendicular ($i^{'}$ = $y^{'}$ or $z^{'}$) component of the bond polarizability. $\alpha^{0}_{i^{'}}$, $\alpha^{1}_{i^{'}}$ and $\alpha^{2}_{i^{'}}$ are constants for a particular bond type.

To determine the values of the constants for the intramolecular interactions, we fit the  model parameters to  the TDDFPT-calculated polarizability tensor data of the monomer H$_2$O. As described in our previous work, this results in essentially perfect agreement between the model and TDDFPT $\pmb{\alpha}$ values.~\cite{GBPM}

To model the polarizability contributed by the intermolecular interactions among  which the intermolecular hydrogen bonds play the key role, we conduct an MD simulation of the (H$_2$O)$_{4}$ cluster at 50~K for 30 ps with a time step of 1.5 fs and then evaluate the $\pmb{\alpha}$ tensor of the cluster along the MD trajectory at the intervals of 3~fs for a total of 10000 structures.
 Since the GBPM model gives essentially perfect agreement with TDDFPT for the water monomer, we decompose the total $\pmb{\alpha}$ tensor according to

\begin{equation}
  \pmb{\alpha}^{\text{total}}=  \pmb{\alpha}^{\text{monomer}} + \pmb{\alpha}^{\text{intermolecular}}
\end{equation}

where we evaluate $\pmb{\alpha}^{\text{monomer}}$ using GBPM and then obtain $\pmb{\alpha}^{\text{intermolecular}}$ by subtracting the GBPM-evaluated $\pmb{\alpha}^{\text{monomer}}$  from the TDDFPT-evaluated $\pmb{\alpha}^{\text{total}}$.

To represent the polarizability contribution of the hydrogen bonds, we follow the BPM formalism and consider each hydrogen bond to contribute parallel and perpendicular bond polarizabilities.  In contrast to the intramolecular O-H bonds, a Taylor series expansion around the equilibrium bond length cannot be used to represent the dependence of $\alpha^{\text{H-bond}}_{\text{parallel}}$ and $\alpha^{\text{H-bond}}_{\text{perpendicular}}$ on the length of the bond ($L^{\text{H-bond}}$) because $\alpha^{\text{H-bond}}_{\text{parallel}}$ and $\alpha^{\text{H-bond}}_{\text{perpendicular}}$ must decay to zero at large $L^{\text{H-bond}}$.
Furthermore, the expression for $\pmb{\alpha}^{\text{intermolecular}}$ must be sufficiently flexible to represent a nonmonotonic function.  We therefore chose to represent the dependence of  $\alpha^{\text{H-bond}}_{\text{parallel}}$ and $\alpha^{\text{H-bond}}_{\text{perpendicular}}$ on $L^{\text{H-bond}}$ as  

\begin{equation}
\alpha^{k}_{i^{'}} = \sum_{j=1}^{m} A_{i^{'}}^{j} \frac{e^{-L}}{L^{j}}  
\label{eq5}
\end{equation}
here, $L$ is the distance between two atoms contributing to the intermolecular interaction, $m$ is the number of the parameters of the series and $A_{i^{'}}^{j}$  are the adjustable  parameters to be determined by fitting to the TDDFPT water cluster data.

We also use the functional form of Eq.~\ref{eq5} to model the effects of the intermolecular H-H and O-O interactions.  We refer to the model including both intramolecular and intermolecular terms as the GBPMh model.

We then fit the parameters in Eq. ~\ref{eq5} to the $\pmb{\alpha}^{\text{intermolecular}}$ values along the MD simulation trajectory of the (H$_{2}$O)$_{4}$ water cluster and use the resultant model to obtain the values of the $\pmb{\alpha}$ along the MD simulation trajectory.

After the polarizability trajectory is obtained,  the Raman spectra can be calculated using the Fourier transform of the auto-correlation function of the electronic polarizability.
In this study,  we calculated the Raman spectra based on the anisotropic part ($\pmb{\beta}$) of the total polarizability ($\pmb{\alpha}$) which is defined as $\pmb{\beta}$ = $\pmb{\alpha}$ - $\frac{1}{3}\textnormal{Tr}(\pmb{\alpha}$)). We choose to consider the time derivative of the anisotropic part of the total polarizability while calculating the Raman spectra using the Fourier transform of the autocorellation functions as expressed in the following formula,
\begin{equation}
R_{\textnormal{aniso}}(\omega) = n_{\textnormal{BE}}(\omega)\int_{-\infty}^{\infty}\textnormal{d}t \;\;\; \textnormal{cos}(\omega t) \;\; \textnormal{Tr}\langle \pmb{\dot{\beta}}(0).\pmb{\dot{\beta}}(t)\rangle  
\label{eq6}
\end{equation}
here, $n_{\textnormal{BE}}(\omega)$ (1 - e$^{-\frac{\hbar \omega}{k_{B}T}}$) is the Bose-Einstein (BE) factor.

In order to determine the unknown optimal values of the adjustable parameters of the polarizability model, we first considered the H$_{2}$O molecule and extracted the GBPM parameters considering the TDDFPT-calculated electronic polarizability trajectory at 100 K, 300 K and 500 K, respectively. To extract the unknown parameters of  Eq.~\ref{eq5} for the GBPMh model, we considered the TDDFPT-calculated polarizability trajectory of (H$_{2}$O)$_{4}$ at 50 K. After extracting the unknown parameters related to GBPM and GBPMh from H$_{2}$O and (H$_{2}$O)$_{4}$, respectively, we applied the model on other water clusters at 50 K, i.e. (H$_{2}$O)$_{r}$ ($r$ = 2, 3, 5 and 6).
We also applied the model to calculate Raman spectra of water clusters at higher temperature and liquid water. All of the model-calculated results are compared either with the TDDFPT or experimental results.

In this work, the TDDFPT-calculated~\cite{MALCIOGLU20111744,GE20142080} electronic polarizability trajectories were obtained for the corresponding $ab$ $initio$ Born-Oppenheimer molecular dynamics trajectories  generated  using the  Quantum ESPRESSO code.~\cite{QE} We used GBRV ultrasoft pseudopotentials~\cite{GARRITY2014446} with an energy cut-off of 50 Ry and the Perdew-Burke-Ernzerhof (PBE) exchange-correlation functional.~\cite{PBE} For the GBPM, the equilibrium structure of H$_{2}$O was obtained after DFT optimization. We considered up to 2$nd$ order of the Taylor series expansion to extract the unknown parameters in the GBPM part. This will introduce twelve unknown parameters in total and six from each type of bond in H$_{2}$O. For the GBPMh, we considered up to $m$ = 7 in the expansion of Eq.~\ref{eq5}, which will introduce forty two parameters in total due to O$\cdots$H, O$\cdots$O, and H$\cdots$H bonds for the parallel and perpendicular components.  

For all of  the clusters including the monomer of H$_{2}$O, $ab$ $initio$ molecular dynamics were performed for a time-step of 0.0015 ps for the time period of 30 ps. The temperatures of the dynamics were set to 100, 300 and 500 K for H$_{2}$O, 50 K for (H$_{2}$O)$_{r}$ ($r$ = 2, 3, 4, 5 and 6). In addition, to check the transferability of the model in the higher temperature clusters with more anharmonic vibrations, we also considered dynamics of (H$_{2}$O)$_{4}$ at 100 K. To model high-temperature hydrogen bond breaking and forming, we carried out MD simulations for the (H$_2$O)$_{4}$ cluster at 363 K in a 8$\times$8$\times$8~\AA$^3$ unit cell. The TDDFPT-calculated polarizability values were calculated considering the time-step of 0.003 ps for a time period of 30 ps for all the systems considered in this work. We also performed classical molecular dynamics at 300 K in LAMMPS code~\cite{LAMMPS} with the AMOEBA~\cite{AMOEBA1,AMOEBA2,AMOEBA3} force fields to obtain a trajectory of liquid water of a time period of 1 ns with a time step of 0.2 fs. For the AMOEBA simulation, we considered 216 water molecules in a box with the size of 18.64 \AA$^{3}$. The Raman spectra of liquid water was calculated based on 100 ps time trajectory with a time step of 0.003 ps.

\section{Results}
\begin{figure}

\centering
\includegraphics[width = 16.2 cm,angle =0]{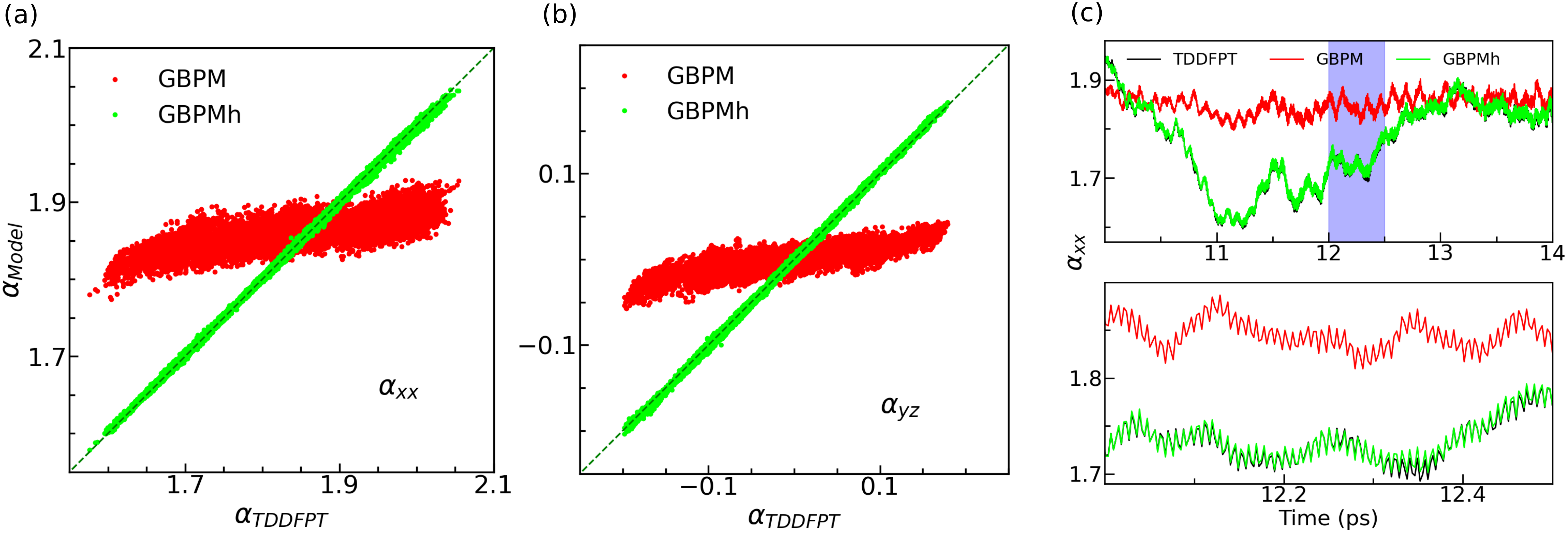}
\caption{(H$_{2}$O)$_{4}$: $\alpha_{Model}$ using GBPM (red) and GBPMh (green) vs $\alpha_{TDDFPT}$ for the component $xx$ and $yz$ in (a) and (b), respectively. Green dotted line represents $y$ = $x$. (c) $\alpha_{xx}$-trajectory calculated using TDDFPT (black), GBPM (red), GBPMh (green) (top panel). Magnified portion of the trajectory as marked using blue color in the top panel is shown in bottom panel.}
\label{Fig2} 
\end{figure}

After extracting the GBPM parameters from H$_{2}$O and intermolecular parameters using Eq.~\ref{eq5} from (H$_{2}$O)$_{4}$, we first test the accuracy of the GBPM and GBPMh models on (H$_{2}$O)$_{4}$.  Figs.~\ref{Fig2} (a) and (b) shows the calculated values of $\alpha_{xx}$ and $\alpha_{yz}$ of (H$_{2}$O)$_{4}$ at 50 K using TDDFPT and model for both GBPM and GBPMh. The  GBPM data (see red points in Figs.~\ref{Fig2} (a) and (b)) show large scatter and strong deviation from the $y$ = $x$ line. This is due to the fact that while GBPM correctly models the intramolecular plarizability of the H$_{2}$O molecules, it does not have the information about the polarizability contributions due to the intermolecular interactions.
By contrast, the GBPMh results for both $\alpha_{xx}$ and $\alpha_{yz}$ (see green points in Figs.~\ref{Fig2} (a) and (b)) show excellent agreement with the TDDFPT results, with all $\alpha$ values showing  only small scatter around  the $y$ = $x$ line.
We then compare the actual polarizability trajectories of (H$_{2}$O)$_{4}$ at 50 K calculated using TDDFPT, GBPM and GBPMh in Fig.~\ref{Fig2} (c), showing  sample low-frequency (top panel) and high-frequency (bottom panel) fluctuations in the selected time range. Clearly, both GBPM and GBPMh models reproduce the high-frequency TDDFPT polarizability fluctuations. This agreement is due to the fact that the high-frequency polarizability fluctuations are due to the vibrations of the internal O-H bonds of the  H$_{2}$O molecules for which the GBPM model shows essentially perfect accuracy. However, GBPM does not contain any information related to the low-frequency polarizability fluctuations arising from the interactions between two or more H$_{2}$O molecules. This makes the overall GBPM trajectory inaccurate, in contrast to the highly accurate GBPMh trajectory. The differences between the models are further reflected in the calculated anisotropic Raman spectra.

\begin{figure}
\centering
\includegraphics[width = 16.2 cm,angle =0]{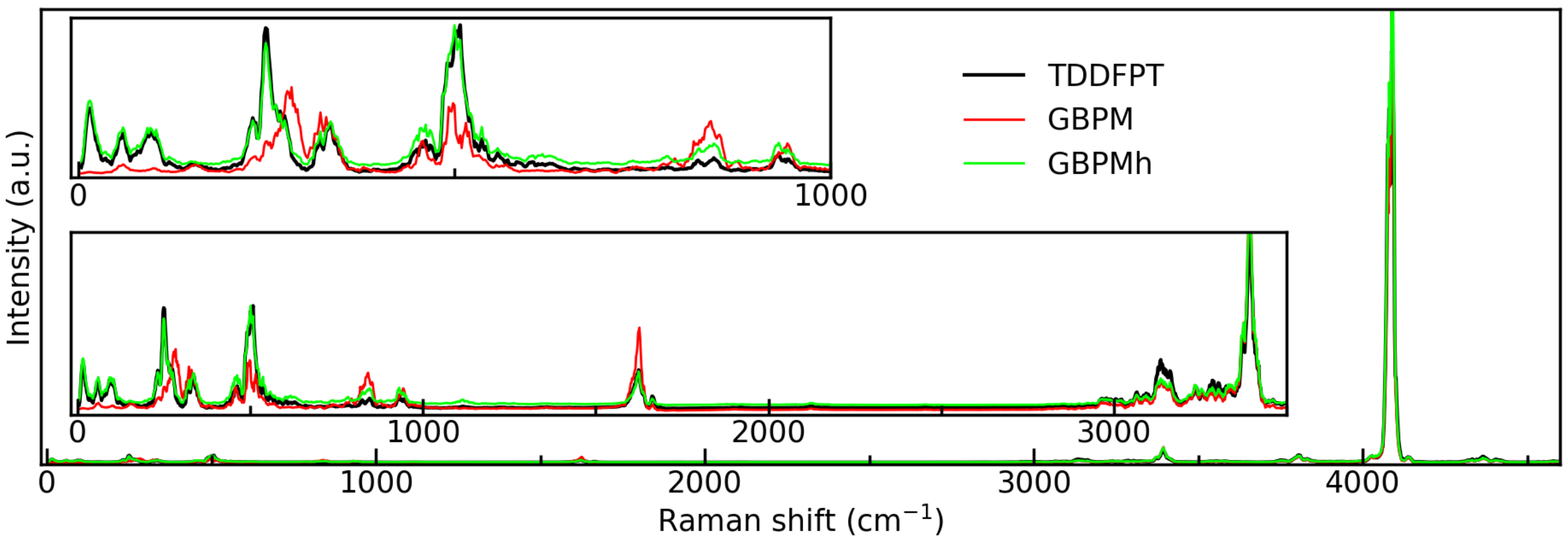}
\caption{(H$_{2}$O)$_{4}$: Raman spectra calculated using TDDFPT (black), GBPM (red), GBPMh (green). Insets show the magnified view of the selected regions.}
\label{Fig3} 
\end{figure}

Fig.~\ref{Fig3} compares the anisotropic Raman spectrum of (H$_{2}$O)$_{4}$ calculated by TDDFPT with the spectra obtained using the GBPM and GBPMh models. As expected, the peak positions and intensities of the TDDFPT
spectrum are well reproduced by GBPM for frequencies above 2500 cm$^{-1}$ where the spectrum is due to the intermolecular O-H stretch vibrations.
However, for frequencies below 2500 cm$^{-1}$, the GBPM results deviates from the TDDFPT results. Interestingly, we can see the bending mode Raman peak which was perfect in the GBPM monomer H$_{2}$O spectrum,~\cite{GBPM} deviates from the TDDFPT peak for the (H$_2$O)$_{4}$ cluster.
Additionally discrepancies between the peak intensities are observed for the peaks at 500 cm$^{-1}$ and  250 cm$^{-1}$.
The greatest disagreement between the TDDPFT and  GBPM spectra is observed for the spectral region below 200 cm$^{-1}$ where the GBPM spectrum is almost flat while the TDDFPT spectrum shows strong peaks.
In contrast to GBPM, GBPMh reproduces all of the peak DFPT peak positions accurately and also reproduces almost exactly the intensities of the peaks with only a small discrepancy at approximately 900 cm$^{-1}$.
Therefore, based on these results, we can divide the spectrum into three parts: (1) the high wave-number peaks contributed by the intramolecular interactions only, (2) the lowest wave-number peaks contributed by the intermolecular interactions only, (3) the intermediate wave-number regions contributed by the coupling between the intramolecular and intermolecular interactions. The mismatch of the Raman intensities at 900 cm$^{-1}$  may be due to the coupling of the intramolecular and intermolecular interactions.

\begin{figure}
\centering
\includegraphics[width = 16.2 cm,angle =0]{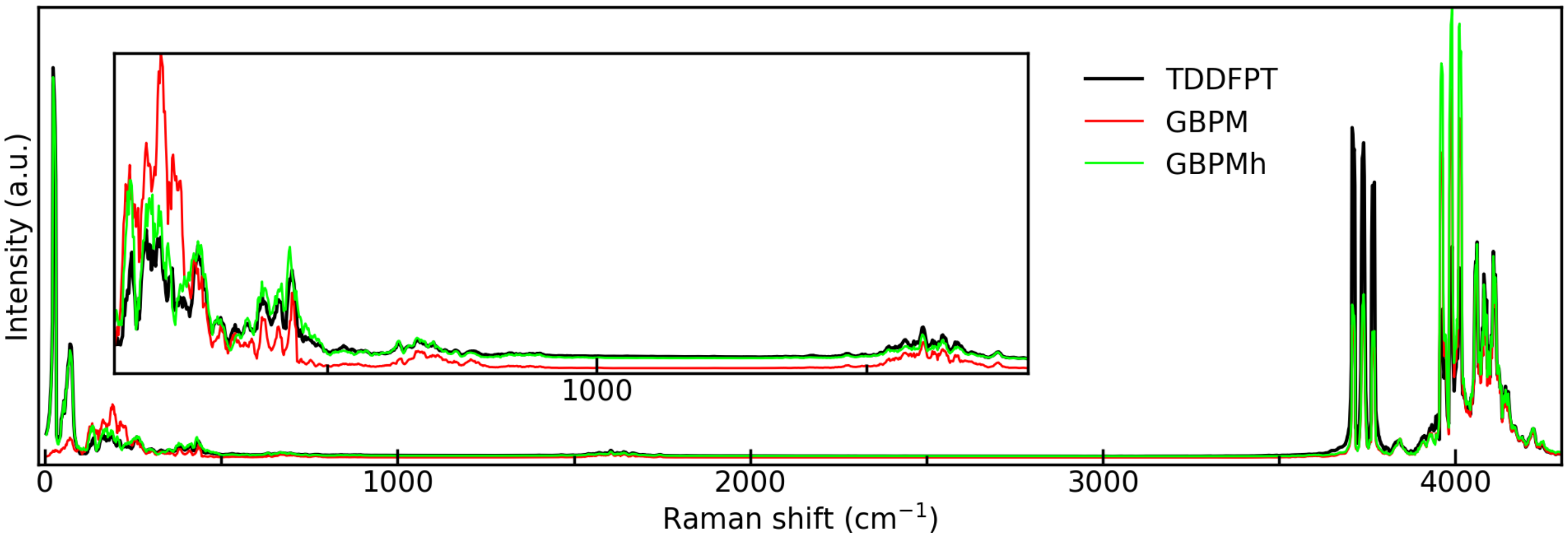}
\caption{(H$_{2}$O)$_{2}$: Raman spectra calculated using TDDFPT (black), GBPM (red), GBPMh (green). Inset shows the magnified view of the selected regions.}
\label{Fig4} 
\end{figure}

\begin{figure}
\centering
\includegraphics[width = 16.2 cm,angle =0]{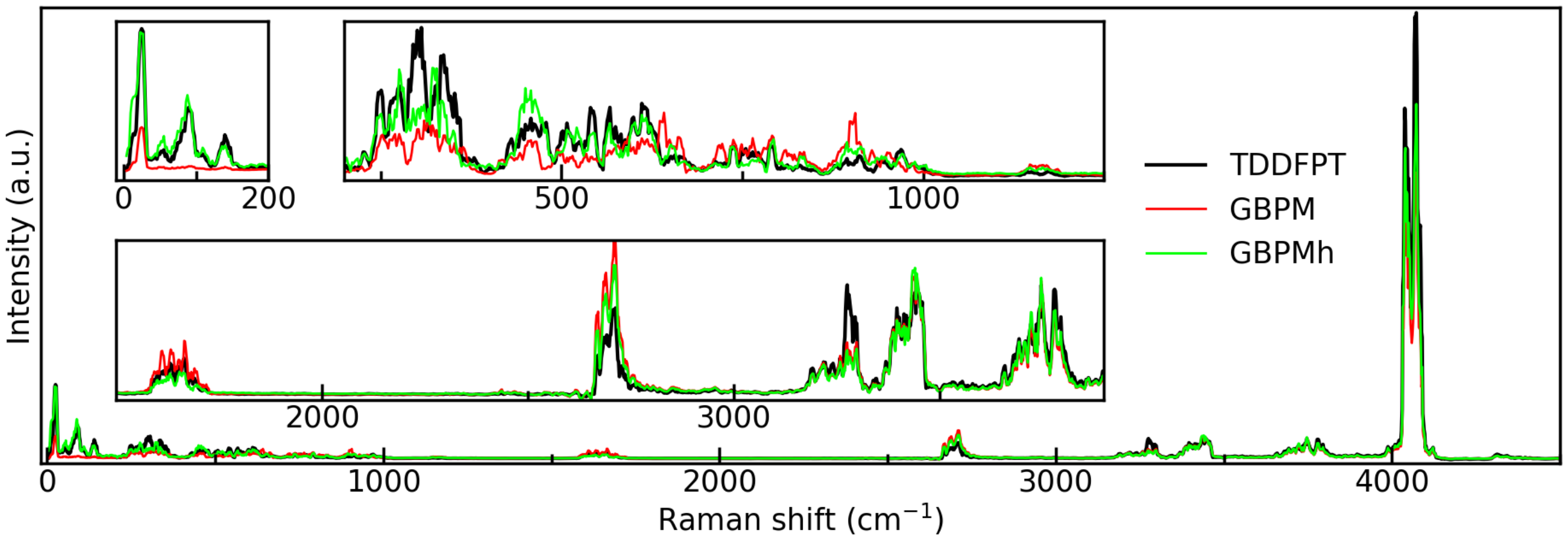}
\caption{(H$_{2}$O)$_{6}$: Raman spectra calculated using TDDFPT (black), GBPM (red), GBPMh (green). Insets show the magnified view of the selected regions.}
\label{Fig5} 
\end{figure}

Next, we use the parameterized GBPM and GBPMh models to calculate the spectra of other H$_{2}$O clusters in order to test the transferability of the model. Figs.~\ref{Fig4} and ~\ref{Fig5} show the  TDDFPT, GBPM and GBPMh Raman spectra for (H$_{2}$O)$_{2}$ and (H$_{2}$O)$_{6}$, respectively.
The results are similar to those obtained for the (H$_{2}$O)$_{4}$ spectra for both GBPM and GBPMh-calculated Raman spectra of these clusters.
The peak positions of the Raman spectra calculated using both  models agree well with the TDDFPT-calculated spectra in the high wave-number region above 1500 cm$^{-1}$. However,  the peak intensities in that region show some differences from the TDDFPT spectrum. While GBPMh does not show a perfect agreement with the TDDFPT spectrum in the spectral region below 1500 cm$^{-1}$, its accuracy in reproducing the TDDFPT spectra is significantly higher.
In particular, the lowest wave-number peaks below 200 cm$^{-1}$ are missed completely by the GBPM and are recovered perfectly by GBPMh for both the clusters.
Further examination of the transferability of GBPMh by comparing the GBPMh and TDDFPT Raman spectra for the (H$_{2}$O)$_{3}$ and (H$_{2}$O)$_{5}$ water clusters showed results similar to those obtained for the (H$_{2}$O)$_{2}$ and (H$_{2}$O)$_{6}$ clusters (see Supporting Information).

We then applied the model to calculate the Raman spectra of water clusters (H$_{2}$O)$_{4}$ at higher temperatures of 100 K and 363 K.  The oscillations of the clusters at 50~K are relatively weak so that the polarizability trajectory obtained by TDDFPT from MD simulations at 50 K samples only a small, close-to-ground-state part of the phase space  of possible water molecule configurations. Therefore, a model parameterized using this data will only be able to reproduce the spectra at higher trajectories if it correctly captures the basic physics underlying the polarizability changes and their functional forms over a wide range. By contrast, a fit with a  functional form that is correct only in the narrow range of the configurations included in the parameterization database will show strong deviations from DFT results for configurations sufficiently different from those in the parameterization database, and will show spectra that are different from the TDDFPT spectra. For the (H$_{2}$O)$_{4}$ cluster at 100 K, the Raman spectra obtained by GBPMh show almost perfect agreement with the TDDFPT spectra. However, since the phase space explored by the short MD trajectory at 100 K is fairly similar to that explored at 50 K as shown by the plot of the hydrogen-bond distances shown in Fig. S1 (see Supporting Information), it is expected that good agreement between the model and TDDFPT spectra will be obtained at 100 K.

Therefore, we then perform a more stringent test and compare the Raman spectra obtained for 4H$_{2}$O at 363 K. As shown in Fig. S1 of Supporting Information, at this temperature the hydrogen bond distances show much larger fluctuations. Hydrogen bonds in the ranges of 1.7-2.5~\AA, 2.5-3.2~\AA~and 3.2-4.0~\AA~are characterized as strong mostly covalent, moderate  mostly electrostatic and weak electrostatic, respectively. Thus, Fig. S1 in Supporting Information shows that the hydrogen bonds in this cluster dynamically change from moderate to weak in some cases, explore the entire range of bond lengths of strong covalent hydrogen bonds, and also break and reform. Thus, the configurations encountered in this trajectory are quite different with regard to hydrogen bonding than the configurations obtained from 50~K MD simulations and used for model parameterization,  and are somewhat similar to the configurations encountered in liquid water where hydrogen bonds are continuously formed and broken. As shown in Fig. S5 of Supporting Information, the GBPMh model reproduces the TDDFPT spectrum of 4H$_{2}$O at 373 K very well, with  some differences in the relative intensities of the high-frequency peaks at 3500~cm$^{-1}$ and 4000~cm$^{-1}$. This  suggests that the functional forms for the intermolecular interactions derived in the model parameterization are physically reasonable.



Next, we applied the model to calculate Raman spectra of liquid water where continuous hydrogen bond breaking occurs during the dynamics. Full DFT calculations of MD trajectory with a large number (216) of molecules are too computationally expensive, necessitating the use of atomistic potentials for MD simulations. Here we use the AMOEBA model to obtain the MD trajectory. Based on the good model results obtained by GBPMh on 4H$_{2}$O at 373 K, the water molecule configurations encountered in liquid water should be accurately treated by GBPMh.  However, since  the polarizability fluctuations depend on the MD trajectory which is obtained using the AMOEBA model, some error relative to the experimental spectra may be introduced by the use of the AMOEBA potential, which is known to provide a good but not perfect representation of liquid water.  

\begin{figure}
\centering
\includegraphics[width = 8.2 cm,angle =0]{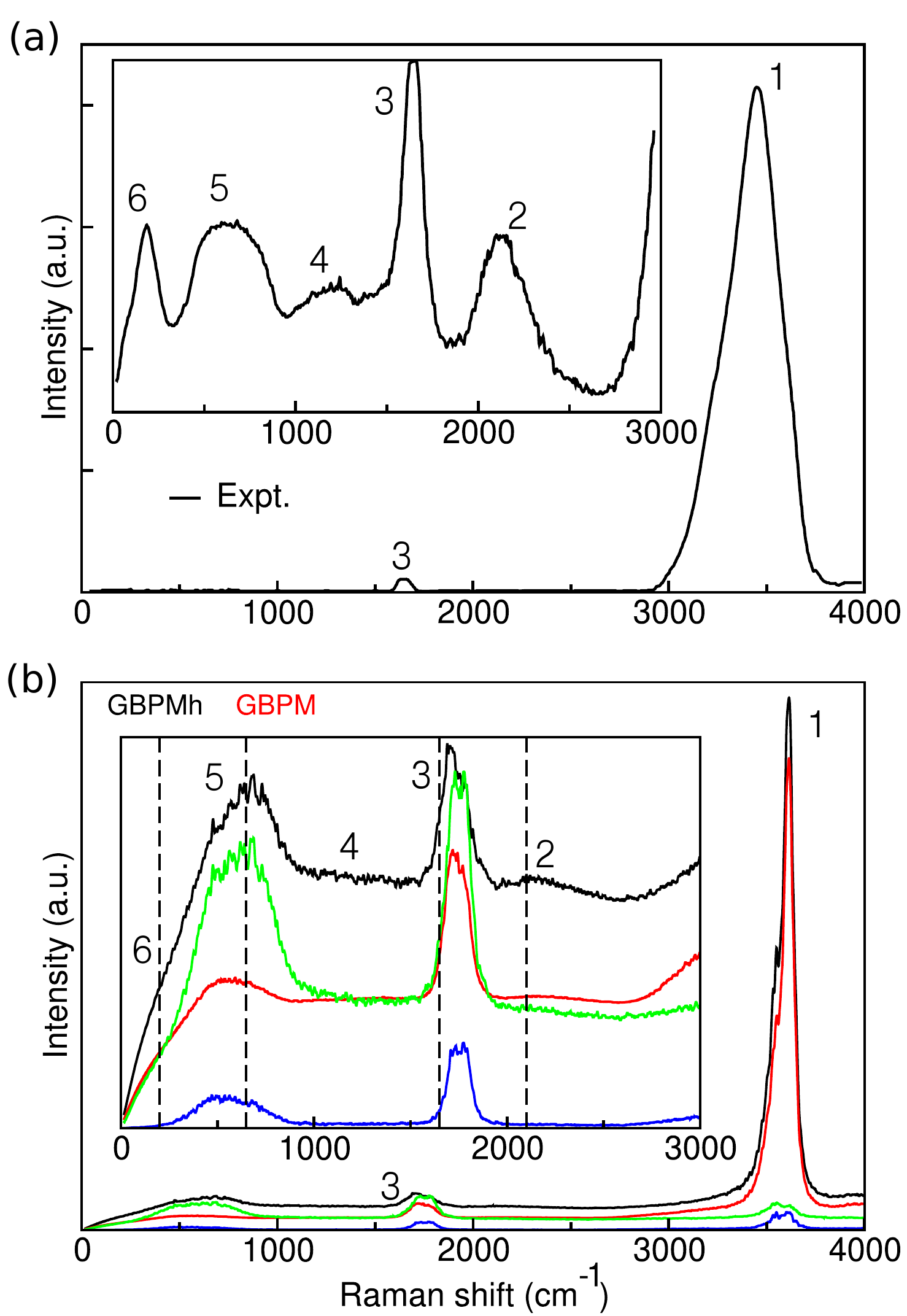}
\caption{Anisotropic Raman spectra of liquid water using (a) experiment (Expt)~\cite{expt_Raman_aniso}, (b) GBPM (red) and GBPMh (black). Spectra based on  intermolecular-intermolecular (green) and intermolecular-intramolecular (blue) part in the auto-correlation function. Inset shows the magnified view of the low wave-number regions. Dotted vertical lines in the inset shows the experimentally measured position of the corresponding peaks.  }
\label{Fig6} 
\end{figure}

Fig.~\ref{Fig6} show the comparison of the calculated Raman spectra using GBPM and GBPMh along with the experimentally measured Raman spectra. The experimental anisotropic Raman spectrum of water contains six peaks (peaks 1, 2, 3, 4, 5 and 6 at 3450, 2100, 1650, 1200, 650 and 200 cm$^{-1}$, respectively) as marked in the Fig.~\ref{Fig6} (a)).

Peak 1  originates from the OH bond stretching mode and both GBPM and GBPMh models can reproduce the high intensity and approximate position of the experimental peak. For both GBPM and GBPMh, the position of peak 1 blueshifted by 160 cm$^{-1}$ compared to the experimental peak 1. This discrepancy is associated with the lack of the quantum mechanical effects for the H atoms in the classical modeling of liquid water dynamics by the AMOEBA force field. Additionally, the peak is more narrow than the experimental peak; since the high-frequency OH stretch is reproduced very well by GBPM and GPBMh models in gas-phase molecules and clusters, the smaller width of the simulation peak is due to our use of AMOEBA to obtain the MD trajectories. A similarly more narrow OH stretch peak is observed for in the spectra obtained from gas-phase and cluster water MD simulations performed using AMOEBA, as shown in Fig. S9 in Supporting Information.

For peak 2, the experimental spectrum and both model spectra show  peak at approximately 2120 cm$^{-1}$; however, the peak intensity is much lower in the calculated GBPMh spectrum than in the experimental spectrum, and the peak is extremely faint in the GBPM spectrum, thus demonstrating significant improvement by GBPMh due to the inclusion of the intermolecular interactions in the polarizability model. The inaccuracy of the peak intensity can be attributed to the deficiency of the AMOEBA classical modeling of the Fermi resonance between the bend overtone and the OH stretching mode.

Peak 3 is associated with the bending mode of the water molecules and is reproduced well by both GBPM and GBPMh models with similar peak intensities. However, the peak position calculated using GBPMh is redshifted by 25 cm$^{-1}$ from the GBPM peak at 1730 cm$^{-1}$ toward the position of the experimental peak, showing better accuracy of the GBPMh results.

Peak 4 in the experimental Raman spectrum is missing in both GBPM and GBPMh spectra. This peak is also absent in the Raman spectrum obtained by the deep neural network model of Sommers et al.~\cite{car}, suggesting that the lack of this peak may be  associated with the classical nature of the simulations or is due to the shortcomings of the AMOEBA potential and DFT calculations of the liquid water dynamics trajectories. 

An examination of the peak 5 shows that this peak is reproduced by both GBPM and GBPMh. However, the peak intensity is much higher for the GBPMh-calculated spectrum than in the GBPM spectrum. It is also observed that  peak 5 in the GBPMh-calculated spectrum is at located 680 cm$^{-1}$ showing a distinct blueshift in comparison to the peak position at 550 cm$^{-1}$ in the GBPM-calculated spectrum, again demonstrating the better agreement of the GBPMh results with the experiment.

Peak 6 appears at 200 cm$^{-1}$ in the experimental spectrum and has been assigned to the vibration of the hydrogen bonds. In contrast to the experimental spectrum, GBPM and GBPMh spectra do not show a separate peak at 200 cm$^{-1}$; however, a close inspection of these spectra shows that they both exhibit a shoulder at 200 cm$^{-1}$, such that the broad peak from 0 to 1000 cm$^{-1}$ can be deconvoluted into two peaks centered at 200 cm$^{-1}$  and 500 cm$^{-1}$.  The shoulder is faint in the GBPM spectrum and more clear in the GBPMh spectrum, again demonstrating the better agreement of the GBPMh spectrum with the experimental Raman spectrum. 

We believe that this discrepancy between GBPMh and experimental spectrum in the peak intensity is not related to the GBPMh model but rather is due to the deviation of the MD trajectory of the AMOEBA force field from the real experimental trajectory.  A similar shoulder-like feature related to the hydrogen bond vibrations is also observed in the infrared spectra of liquid water  calculated from MD trajectories obtained using the AMOEBA potential in  a previous work.~\cite{acs.jpcb.2c04454} The imperfect accuracy of the AMOEBA force field is also manifested in the smaller width  of the high-frequency O-H stretch peak (peak 1) of the AMOEBA spectrum compared to the experimental spectrum, as discussed above. 

The analytical formulation of the $\pmb{\alpha}$ tensor in the GBMPh model and its separation of the intra- and inter-molecular contributions enables easy analysis of the total Raman spectrum in terms of the different contributions. Since the $\pmb{\alpha}$ tensor can be written as a sum of the $\pmb{\alpha}^{\text{monomer}}$ and $\pmb{\alpha}^{\text{intermolecular}}$, we can decompose the Raman spectrum into the intramolecular, intermolecular and joint intermolecular-intramolecular contributions, where the intramolecular contribution is the same as the GBPM spectrum.  These spectra are plotted in Fig.~\ref{Fig6} (b)  together with the total GBPMh spectrum obtained from AMOEBA trajectory.  As expected, peak 1 is almost entirely  due to the intramolecular interaction, while peak 2 has contributions from both intramolecular and intermolecular interactions. For peak 3, strong contributions are present from all three interactions, showing that the angle bend is strongly affected by the hydrogen bonding.  For peaks 5 and 6, the intermolecular contribution is the strongest showing that these peaks are in fact due to the vibrations of hydrogen bonds.

\section{Discussion}

Several research directions can be pursued in the further development of the contribution of hydrogen-bonding to polarizability and Raman spectra. First, the functional form used to represent intermolecular interactions can be improved.  While the basis set  of functions that we used in the GBPMh model allows good fit of the DFT polarizability trajectory, it would be better to use a rigorously complete basis set would enable the use of fewer terms. Additionally, while the intermolecular polarizability must decay to zero at large distances it is unclear whether the optimal coefficient for the exponential is 1 as used in our current model. Thus, there is room for improvement in the details of the functional form of intermolecular interactions that will be addressed in future work.
Second, the current model ignores the fact that the oxygen atom in water has lone pairs. It is possible that further improvement in model quality will be obtained by going to a 4-point model of the water molecule  where the O$\cdots$H and O$\cdots$O intermolecular interactions are represented as interactions involving the O lone pairs rather than the position of the O atom.   
Third, the GBPMh model can also be applied to other molecular systems for which hydrogen bonding is important such as alcohols, ketones and acids, enabling modeling of Raman spectroscopy of a wide range of aqueous solutions important in chemical and biological systems. Fourth,  the interactions introduced in the GBPM and GBPMh models can be used for representations of the local environment for use in neural network models.  The use of a representation that is more suitable to the description of the polarizability will likely reduce the size of the neural network necessary to model the polarizability changes during the dynamics.  For example this approach has been used in a recent work on modeling of Raman spectra of biphenyl and malonaldehyde where BPM interactions were used to construct neural networks for polarizability modeling.~\cite{acs.jctc.4c01086}

With regard to applications, the GBPMh  model can be applied to the MD trajectories of water obtained using different atomistic potentials  to test their quality by comparing the Raman spectra obtained from the model dynamics with experimental spectra.  Thus, comparison of Raman spectra to the experimental spectra can become a standard benchmark in the development of more accurate atomistic potentials for water.  Future work should also examine the application of this model to Raman spectra of water in different phases and different conditions (e.g. nanoscale confinement) for atom-level interpretation of experimental Raman spectra. Beyond water, the GBPMh approach can be applied to the simulations of Raman spectra of a wide range of  aqueous solutions in which hydrogen bonding plays a key role and which have been extensively studied by experimental Raman spectroscopy to provide high-resolution interpretation of the experimental spectra in terms of local structure motifs and dynamics.~\cite{jp035267+,1.2732745,molecules29133035}

\begin{figure}
\centering
\includegraphics[width = 10.2 cm,angle =0]{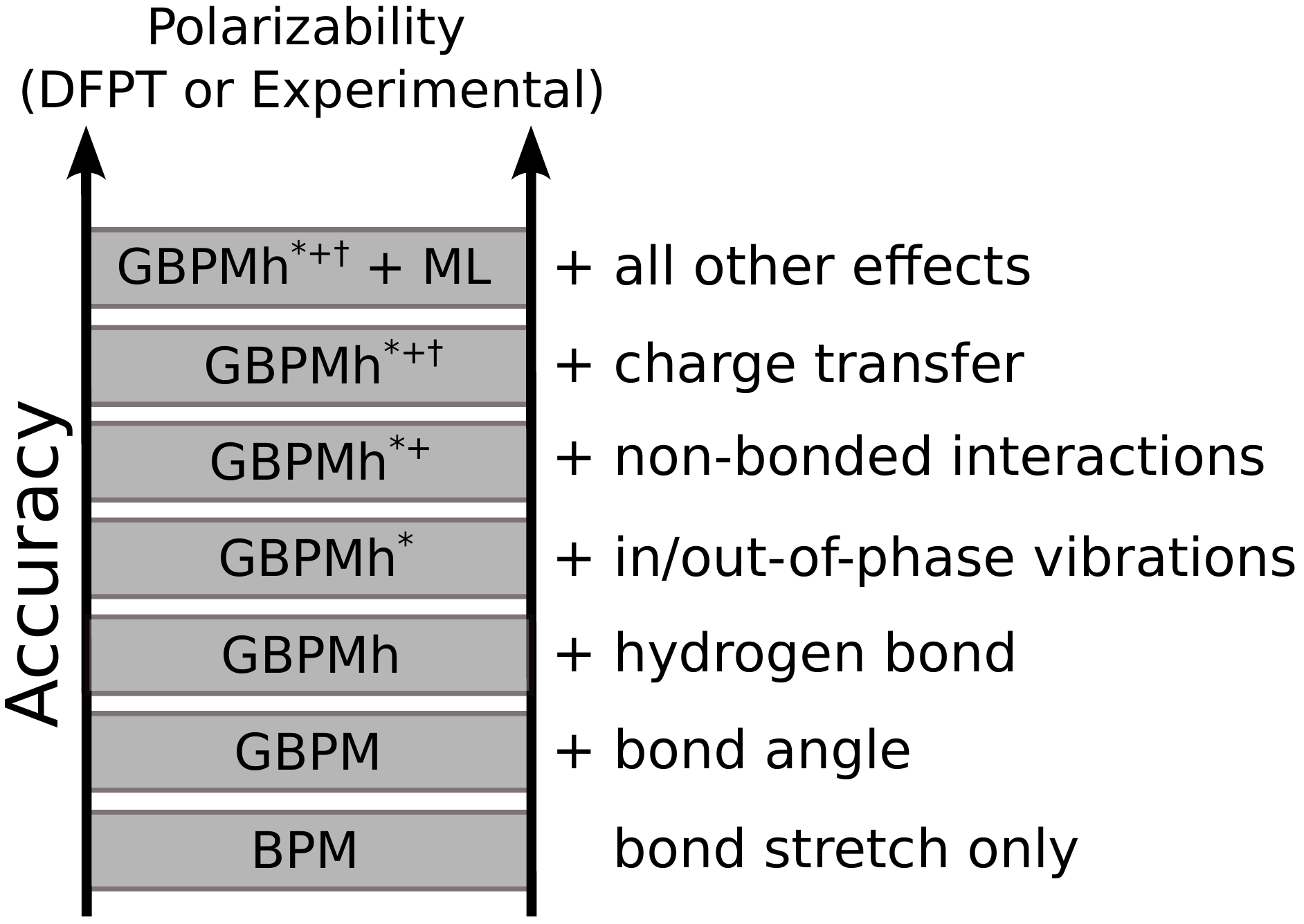}
\caption{Proposed Jacob's ladder towards accurate atomistic polarizability modeling. Accuracy is increasing from lower to upper rungs. Each rung represent the added interaction.}
\label{Fig1} 
\end{figure}

More broadly, the success of GBPM and GBPMh in improving the BPM results to high accuracy through an inclusion of additional interactions suggests that a systematic improvement of the atomistic model can be achieved by including additional physically intuitive effects.  While the exact functional form of the atomistic model of electronic polarizability is not known, the understanding of the chemical bonding and electronic structure of molecules and solids achieved to date enables identification of increasingly complex and non-local interactions  that can affect the polarizability. This is similar to the development of exchange-correlation (XC) functionals in DFT where the exact form of the XC functional is unknown, but this functional is known to obey several constraints and can be systematically improved as famously described by Perdew in the Jacob's ladder of XC functionals.~\cite{1.1390175}  Inspired by this idea, we propose that atomistic models of electronic polarizability can be systematically improved to form a Jacob's ladder of increasingly sophisticated atomistic models of polarizability (see Fig.~\ref{Fig1}) analogous to the Jacob’s ladder of exchange-correlation functional introduced by Perdew.~\cite{1.1390175}

The classic bond polarizability model  provides the foundation for modeling of the dependence of electronic polarizability on atomic coordinates. While highly simplistic, it does reproduce the bond-stretching peaks in Raman spectra. Additionally, for systems with stiff bond angles or with high symmetry, the errors due to the neglect of bond angle effects are small.\cite{GBPM}
Thus, despite its strongly local focus, BPM provides a useful basis for polarizability models and forms the first level of the Jacob’s ladder of the polarizability models.

The second most important effect is the interaction between the positions of the two bonds in space (angle bend).  As we showed in our recent research,~\cite{GBPM} consideration of the quantum mechanical expression for the polarizability  naturally leads to the conclusion that a distance between the atoms typically considered to be non-bonding (e.g., H atoms in H$_{2}$O) also has a strong effect on the polarizability because the HOMO and LUMO orbitals which play the key role in determining the electronic polarizability are often located along this distance.  This leads to the generalized BPM (GBPM) model which shows much better agreement with first-principles polarizability trajectories for molecules and solids and forms the second rung on the polarizability  Jacob’s ladder.

The (relatively) weak intermolecular hydrogen  bonds  are common in chemistry and play the next most important role in the electronic polarizability of aqueous solutions. The vibrations of these  bonds are also highly anharmonic and in liquids these bonds are constantly broken and reformed, making a harmonic approximation focused on the equilibrium bond length inappropriate and necessitating the use of a more complex functional form to describe the changes in the electronic polarizability contributions of these bonds during the dynamics.  Thus, a model that adds terms to GBPM to  accurately simulate the changes in polarizability  due to H-bond dynamics forms the third rung on the polarizability Jacob’s ladder. As shown above, the GBPMh model can accurately reproduce the Raman spectra of water clusters and provides a significant improvement in the results for liquid water.

For solids, the phase of bond vibrations is likely to play the next most important role in electronic polarizability, necessitating the development of a model that can distinguish between long-short-long-short and long-long-short-short bond distance patterns.  Such bond-bond interactions can be taken into account using an atom-centered version of BPM that we have developed that considers the total bond order of each atom in determining the bond polarizability.~\cite{paul2023}
This naturally distinguishes the long-short-long-short (preserving bond order conservation) and long-long-short-short (deviations from bond order conservation) bond distance patterns.  We call a model including this interaction GPBMh$^{*}$, forming the fourth rung on the polarizability Jacob’s ladder. 

The effects of non-bonded interactions such as steric, van der Waals and electrostatic  interactions are next in complexity.  These interactions can be considered as a  field acting on the bond and changing the parameters in the BPM.  Then, we can model the changes in the BPM parameters due to external fields as changes in the $\alpha^k_{i^{'}}$ coefficients in Eq. (3) expressed through  a Taylor series expansion in  the powers of the field. 

This will allow much more accurate modeling of complex systems where non-bonding interactions are important such as water clusters and liquid water. We call this GPBMh$^{*+}$, and this model forms the fifth rung on the polarizability Jacob’s ladder.

To go beyond the  models at the fifth rung, additional accuracy improvements can be obtained by either including other effects such as dynamic charge transfer or by using the variables characterizing different interactions as features in a machine learning model (e.g., SVR or shallow NN) for representing the differences between the true polarizability and that predicted by the fifth-rung model, where the physical relevance of the features used in the ML model will enable accuracy for a computationally inexpensive model.~\cite{shafir2023,acs.jctc.4c01086}
Such a model can be called GBPMh$^{*+\dagger}$.

\section{Conclusions}

We have demonstrated that a simple and accurate  analytical model of electronic  polarizability can be constructed for water clusters by extending the previously reported GBPM model to include intermolecular interactions, most importantly the hydrogen bond O$\cdots$H interactions.  By using a multi-term flexible functional form constrained to obey the correct physical limits  for  this interaction, we obtained excellent agreement between the TDDFPT and model Raman spectra of (H$_2$O)$_{r}$ clusters ($r$ = 2, 3, 4, 5, 6).  Comparison to TDDFPT spectra from MD simulations of clusters at higher temperatures showed that the model has good transferability.
Simulations of liquid water using the AMOEBA model show a strong improvement in the agreement with the experimental liquid water Raman spectrum and reveal the shortcomings of the AMOEBA model in reproducing several spectral features of liquid water.  The low number of parameters (54)
and computational cost of this model make it suitable for use in large MD simulations  of Raman spectra.  The model can also serve as a basis for further development of accurate polarizability modeling for Raman spectroscopy by extending it to other systems, by using the interactions identified to be important as features in machine learning models or by providing a basis for hybrid analytical/NN models where the NN is used to  focus on more complicated interactions, simplifying its training and structure. 

\begin{acknowledgement}
A.P. and I.G. acknowledge the support of the Army Research Office
under Grant W911NF-21-1-0126 and Army/ARL via the
Collaborative for Hierarchical Agile and Responsive Materials (CHARM) under cooperative
agreement W911NF-19-2-0119. A.P. and I.G acknowledge additional support from Israel Science Foundation under Grant 1479/21. 

\end{acknowledgement}

\begin{suppinfo}
The Supporting Information file contains (1) the results of (H$_{2}$O)$_{r}$ ($r$ = 3 and 5 at 50 K, $r$ = 4 at 100 and 373 K), (2) functional form of intermolecular polarizability contributions (parallel and perpendicular) for O$\cdots$H, H$\cdots$H and O$\cdots$O pairs in (H$_{2}$O)$_{4}$ at 50 K, (3) convergence with number of structures of (H$_{2}$O)$_{4}$ at 50 K for GBPMh model parameterization.

\end{suppinfo}

\providecommand{\noopsort}[1]{}\providecommand{\singleletter}[1]{#1}%
\providecommand{\latin}[1]{#1}
\makeatletter
\providecommand{\doi}
  {\begingroup\let\do\@makeother\dospecials
  \catcode`\{=1 \catcode`\}=2 \doi@aux}
\providecommand{\doi@aux}[1]{\endgroup\texttt{#1}}
\makeatother
\providecommand*\mcitethebibliography{\thebibliography}
\csname @ifundefined\endcsname{endmcitethebibliography}
  {\let\endmcitethebibliography\endthebibliography}{}

\end{document}


\section{Results of (H$_{2}$O)$_{r}$ ($r$ = 3 and 5 at 50 K, $r$ = 4 at 100 and 373 K)}

To check the transferability of the GBPMh model based on (H$_{2}$O)$_{4}$ at 50 K, we used the model to calculate the Raman spectra of  (H$_{2}$O)$_{3}$ and (H$_{2}$O)$_{5}$ clusters at 50 K and of the  (H$_{2}$O)$_{4}$ cluster at 100 K. We also considered the simulation of (H$_{2}$O)$_{4}$ placed in a 8$\times$8$\times$8 \AA$^{3}$ supercell at 373 K that  enables us to test our model in a strongly anharmonic environment.

\begin{figure}[h]
    \centering
    \includegraphics[width = 8.2 cm,angle =0]{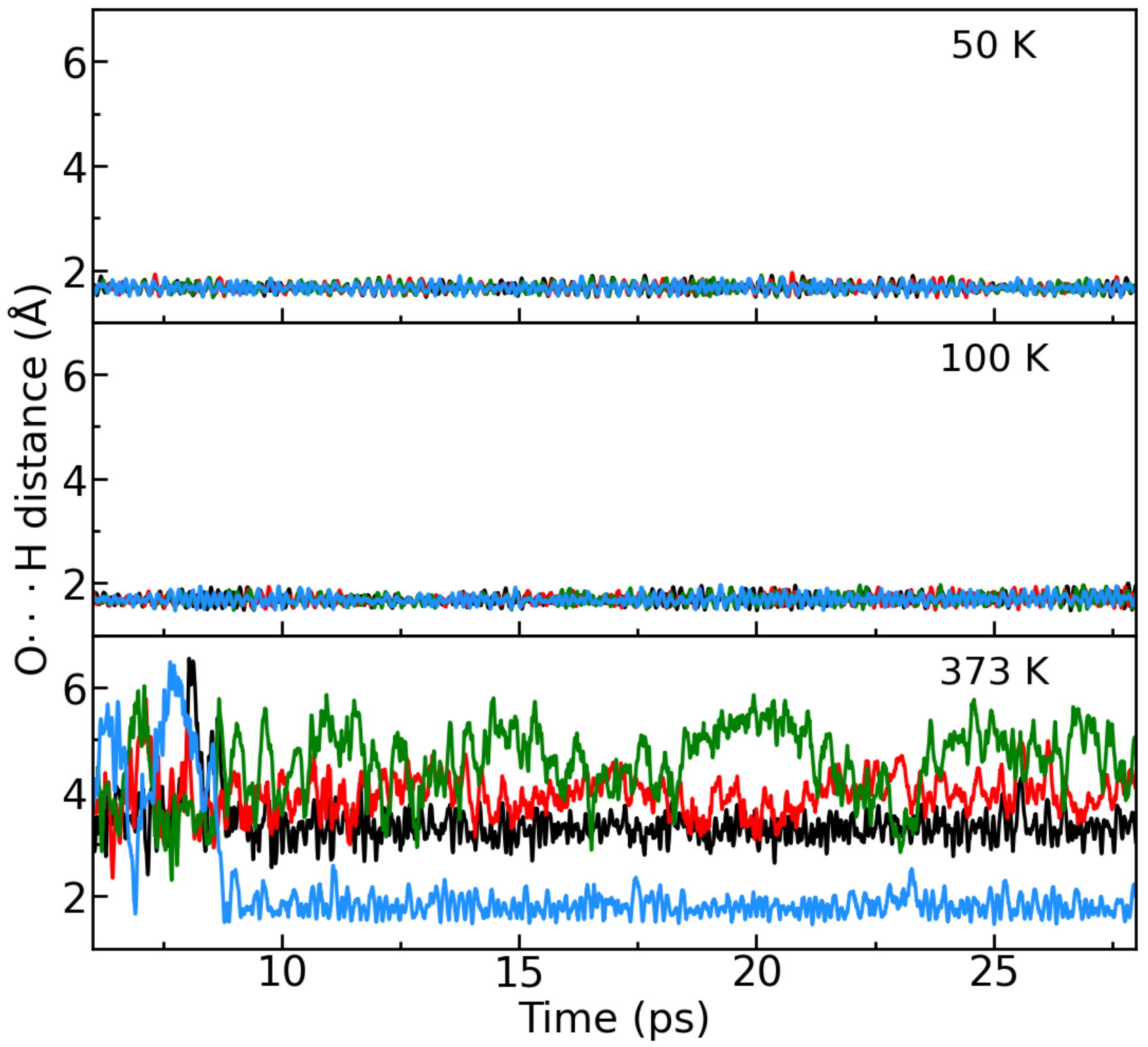}
    \caption{Dynamics of four hydrogen bond (O$\cdot\cdot\cdot$H-pairs) distances of (H$_{2}$O)$_{4}$ at 50 K (top panel) and at 100 K (bottom panel). Dynamics of four intermolecular bond distances (O$\cdot\cdot\cdot$H pairs) of (H$_{2}$O)$_4$ at 373 K. Four different colors (black, red, green and blue) in each panel correspond to four different O$\cdot\cdot\cdot$H-pairs.
    }
    \label{Fig1}
\end{figure}

To show the changes in the dynamics with temperature for  (H$_{2}$O)$_{4}$ clusters, we calculated the O$\cdot\cdot\cdot$H-pair distances corresponding to the four hydrogen bonds as a function of simulation time as shown in Fig.~\ref{Fig1}. For the simulation  at 373 K, it can be expected that breaking and formation of hydrogen bonds will occur during the dynamics due to high temperature which would lead to the dissociation of the cluster for a simulations in a large cell. Due to the small cell, instead of dissociating, the cluster can reform as the H$_2$O molecules are forced to always be in close proximity.   Clearly, the four O$\cdot\cdot\cdot$H-pair distances corresponding to the four hydrogen bonds are stable for the (H$_{2}$O)$_{4}$ cluster at 50 K and 100 K, with  fluctuations around 1.7 \AA~for both temperatures as can be seen in Fig.~\ref{Fig1}. However, at 373 K, the four O$\cdot\cdot\cdot$H-pair distances are showing a large variation between 1.6 and 6.5 \AA~during the dynamics, confirming the breaking and formation of hydrogen bonds, verifying that this simulations explores a large range of hydrogen distances that will also be encountered in simulations of liquid water. 

\begin{figure}[h]
    \centering
    \includegraphics[width = 12.2 cm,angle =0]{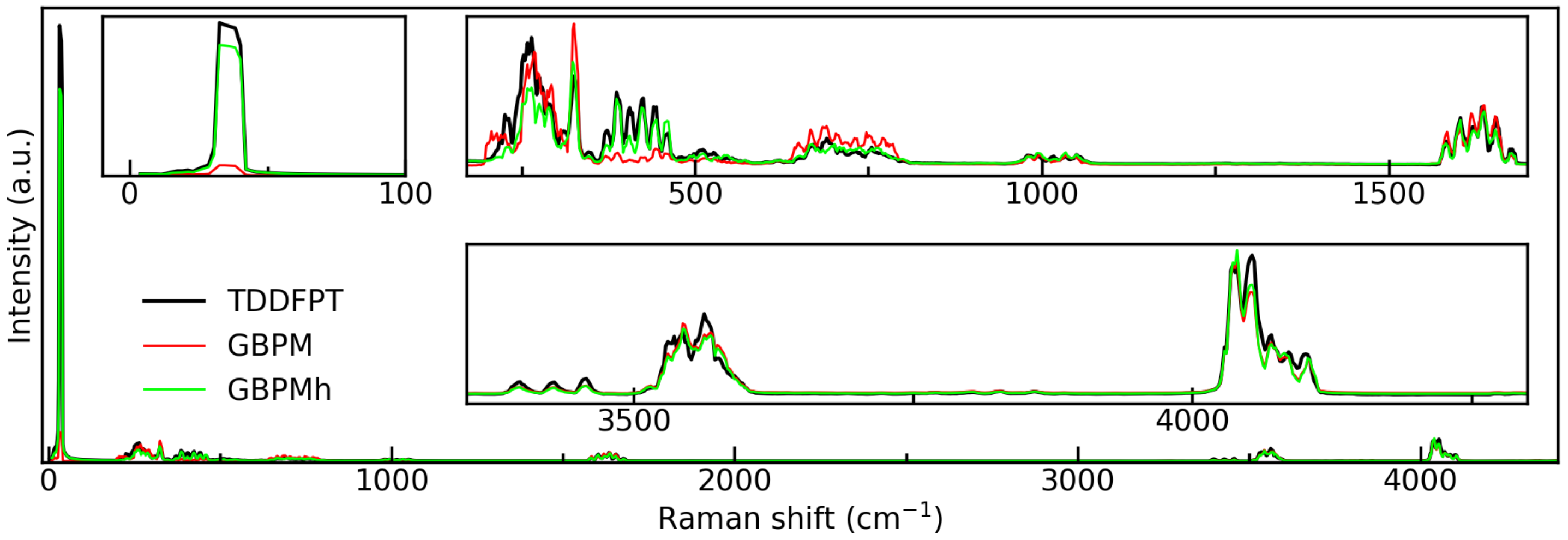}
    \caption{(H$_{2}$O)$_{3}$ at 50 K: Raman spectra calculated using TDDFPT (black), GBPM (red), GBPMh (green). Inset shows the magnified view of the selected regions.}
    \label{Fig2}
\end{figure}

\begin{figure}[h]
    \centering
    \includegraphics[width = 12.2 cm,angle =0]{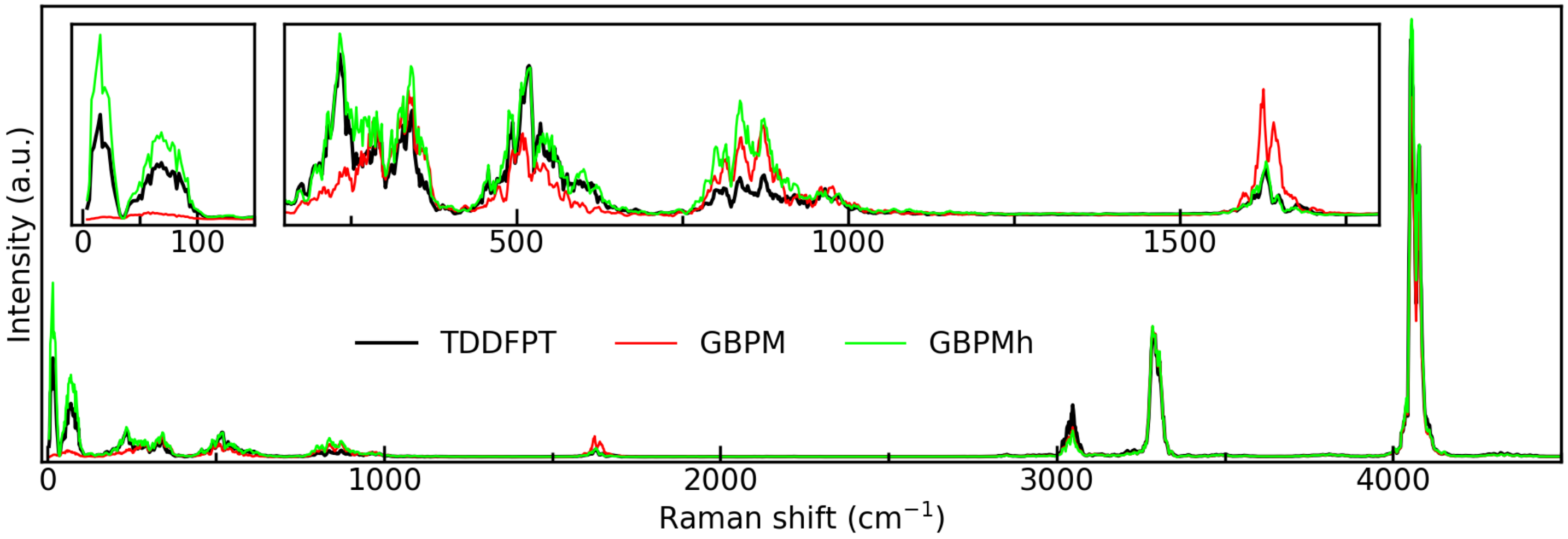}
    \caption{(H$_{2}$O)$_{5}$ at 50 K: Raman spectra calculated using TDDFPT (black), GBPM (red), GBPMh (green). Inset shows the magnified view of the selected regions.}
    \label{Fig3}
\end{figure}

\begin{figure}[h]
    \centering
    \includegraphics[width = 12.2 cm,angle =0]{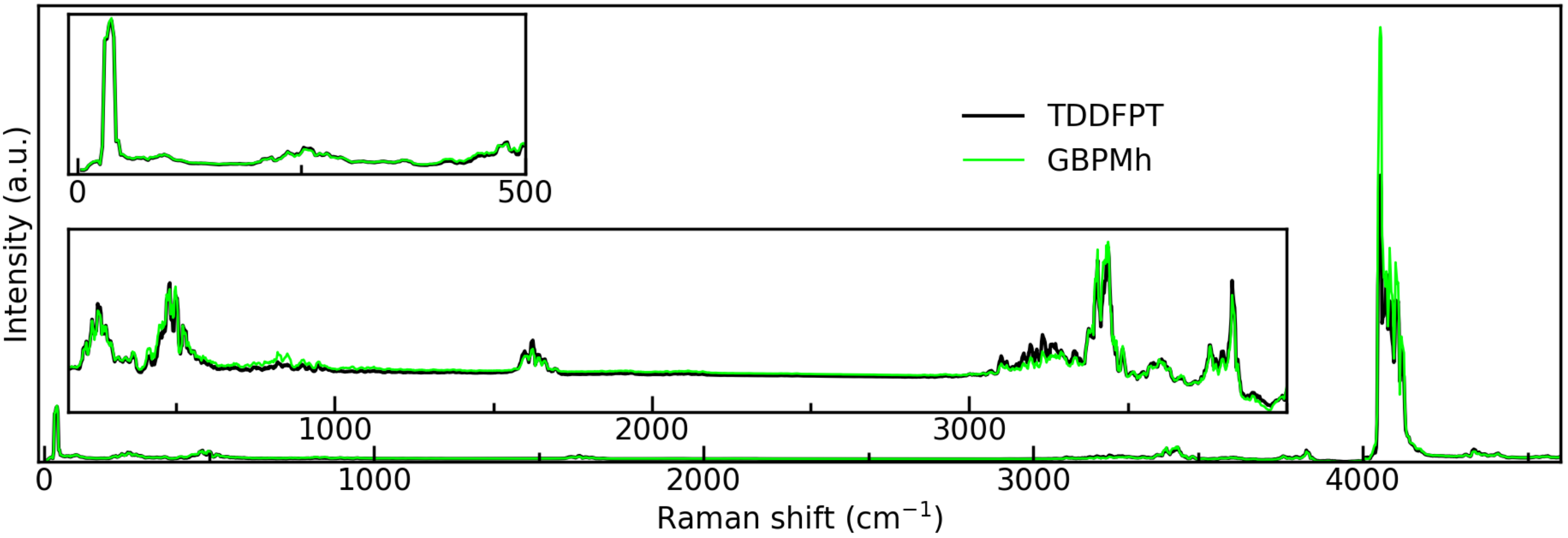}
    \caption{(H$_{2}$O)$_{4}$ at 100 K: Raman spectra calculated using TDDFPT (black), GBPMh (green). Inset shows the magnified view of the selected regions.}
    \label{Fig4}
\end{figure}

\begin{figure}[h]
    \centering
    \includegraphics[width = 12.2 cm,angle =0]{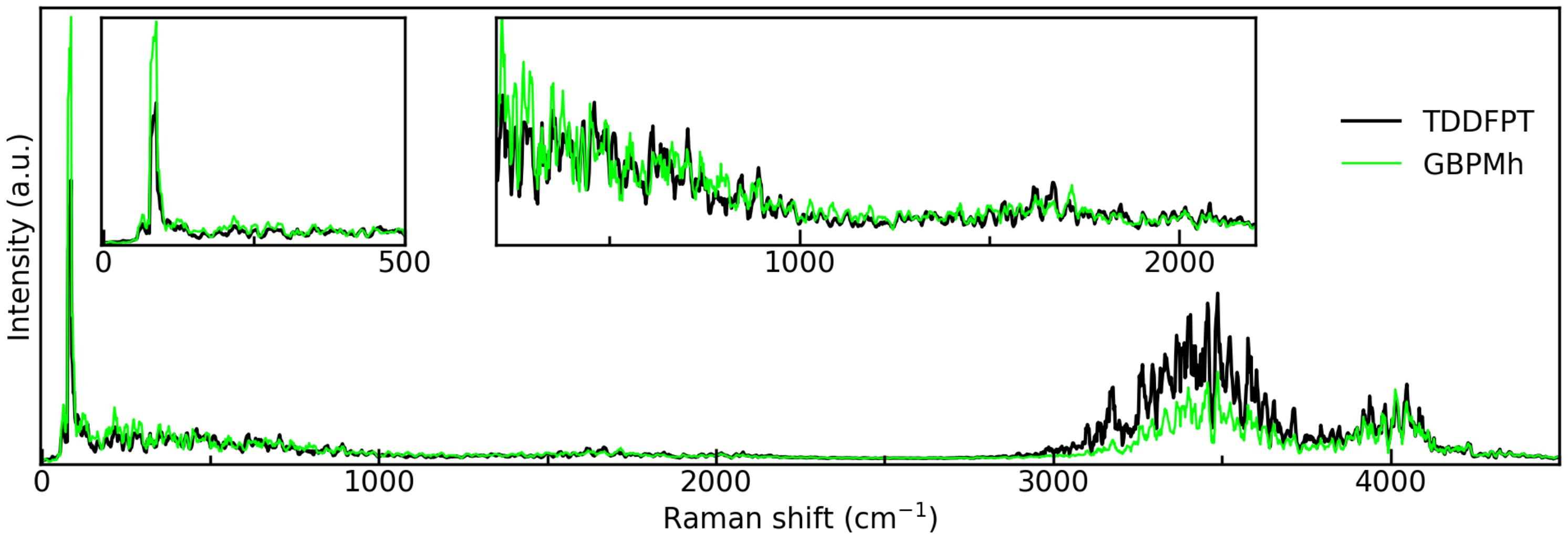}
    \caption{4H$_{2}$O at 373 K: Raman spectra calculated using TDDFPT (black), GBPMh (green). Inset shows the magnified view of the selected regions.}
    \label{Fig5}
\end{figure}

The calculated Raman spectra using GBPM, and the GBPMh model based on (H$_{2}$O)$_{4}$ at 50 K, are compared with the TDDFPT-calculated Raman spectra of (H$_{2}$O)$_{3}$ and (H$_{2}$O)$_{5}$ at 50 K as shown in Fig.~\ref{Fig2} and Fig.~\ref{Fig3}, respectively. Results  similar to those for other clusters are observed for these cases. Though the-high wavenumber peaks are well reproduced by GBPM, the low-wavenumber peaks associated with intermolecular interactions do not reproduce the TDDFPT results well. By contrast, the GBPMh spectra show  very good agreement with the TDDFPT-calculated spectra. To test the performance of GBPMh model based on (H$_{2}$O)$_{4}$ at 50 K in the more anharmonic conditions, we compared the GBPMh with TDDFPT-calculated Raman spectra of (H$_{2}$O)$_{4}$ at 100 K, and 4H$_{2}$O at 373 K, in  Fig.~\ref{Fig4}, and Fig.~\ref{Fig5}, respectively. The comparison of the GBPMh and TDDFPT results shows perfect agreement for the  cluster at 100~K and  very good agreement for the cluster at 373 K. 

\section{Functional form of intermolecular polarizability contributions (parallel and perpendicular) for O$\cdot\cdot\cdot$H, H$\cdot\cdot\cdot$H and O$\cdot\cdot\cdot$O pairs in (H$_{2}$O)$_{4}$ at 50 K}
\begin{figure}[h]
    \centering
    \includegraphics[width = 12.2 cm,angle =0]{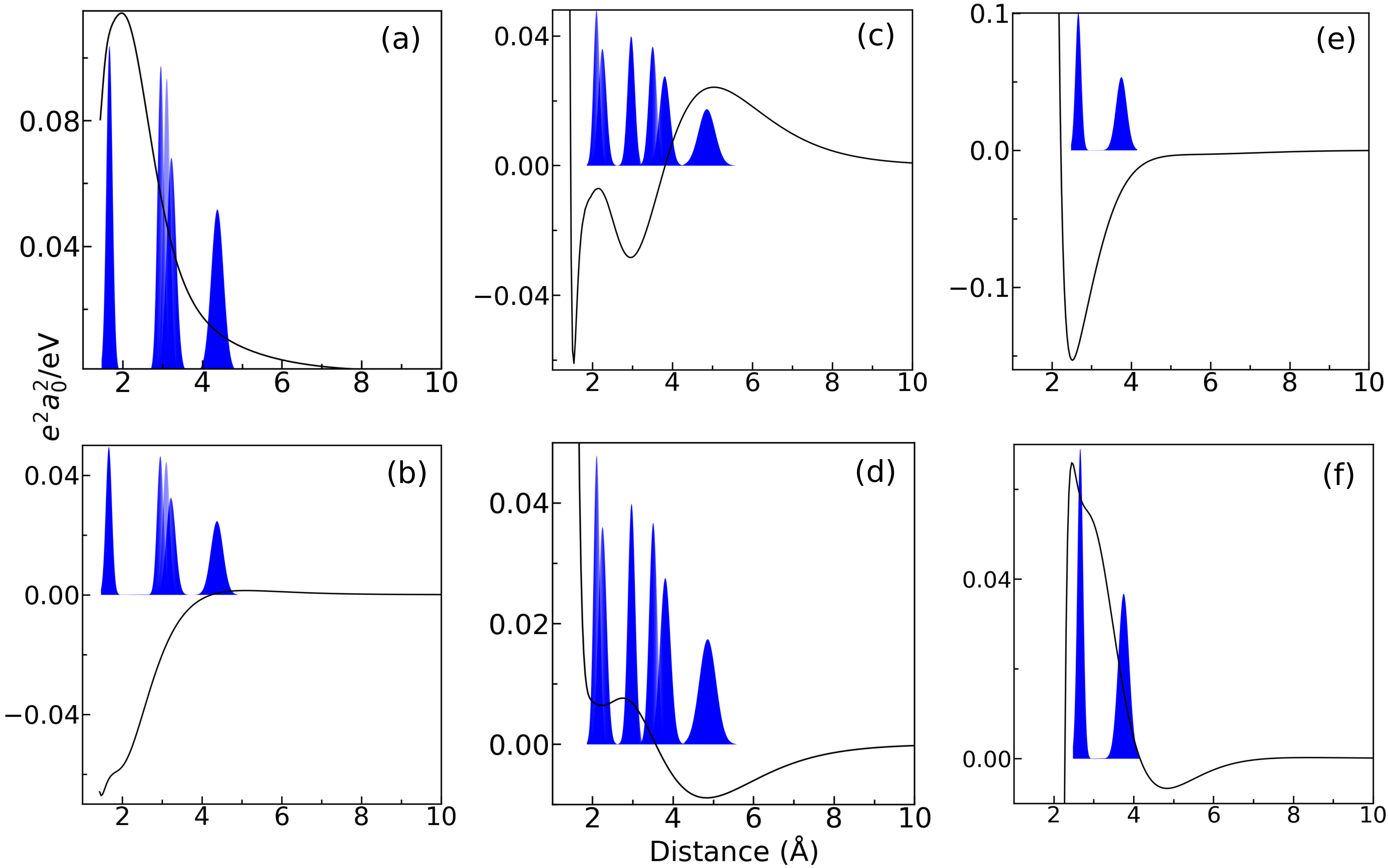}
    \caption{(H$_{2}$O)$_{4}$ at 50 K: Variation of $\alpha$ as a function of O$\cdot\cdot\cdot$H ((a) and (b)), H$\cdot\cdot\cdot$H ((c) and (d)), and O$\cdot\cdot\cdot$O ((e) and (f)) distances for the component parallel (top figures) and perpendicular (bottom figures) to that pair in GBPMh (shown in black line). Distributions of the actual O$\cdot\cdot\cdot$H, H$\cdot\cdot\cdot$H, and O$\cdot\cdot\cdot$O  distances during the dynamics are shown in blue in each plot.}
    \label{Fig6}
\end{figure}

Fig.~\ref{Fig6} shows the dependence of intermolecular polarizability $\alpha$ terms (parallel and perpendicular) of the parameterized GBPMh model for the  O$\cdot\cdot\cdot$H, H$\cdot\cdot\cdot$H and O$\cdot\cdot\cdot$O pairs as a function the corresponding interatomic  distances.
Distributions of the actual O$\cdot\cdot\cdot$H, H$\cdot\cdot\cdot$H, and O$\cdot\cdot\cdot$O  distances during the dynamics are shown in blue in each plot.
The nonmonotonic functional form of the parameterized intermolecular $\alpha$ terms   can be seen in the region below 7 \AA~with decay to zero beyond that distance.  

\section{Convergence with number of structures of (H$_{2}$O)$_{4}$ at 50 K for GBPMh model parameterization}
\begin{figure}[h]
    \centering
    \includegraphics[width = 6.2 cm,angle =0]{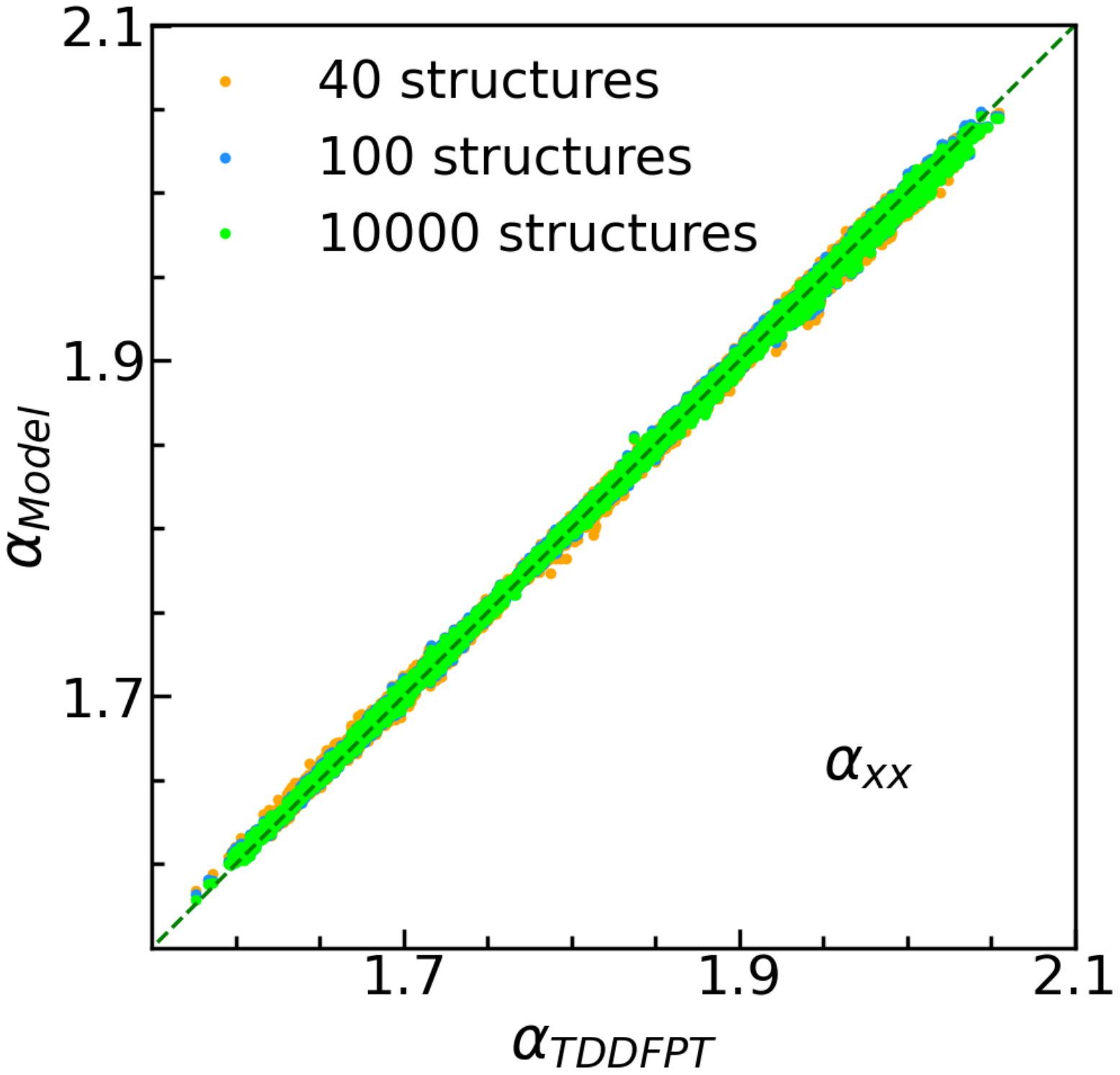}
    \caption{(H$_{2}$O)$_{4}$ at 50 K: Value of $\alpha_{xx}$ of 10000 structures as calculated using GBPMh and TDDFPT. GBPMh is parameterized based on 40 (yellow), 100 (blue), and 10000 (light green) structures. Deep green dotted line represents $y$ = $x$.
    }
    \label{Fig7}
\end{figure}

\begin{figure}[h]
    \centering
    \includegraphics[width = 12.2 cm,angle =0]{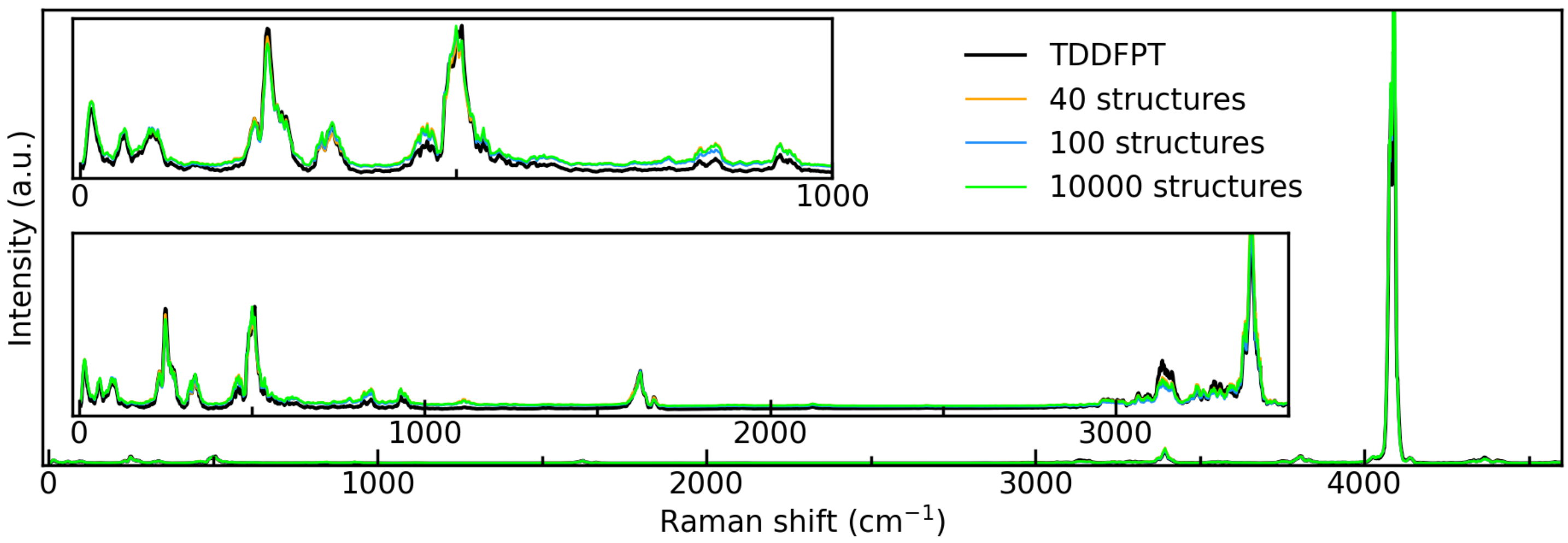}
    \caption{(H$_{2}$O)$_{4}$ at 50 K: Raman spectra using TDDFPT and GBPMh, where GBPMh is parameterized using 40 (yellow), 100 (blue), and 10000 (light green) structures. }
    \label{Fig8}
\end{figure}

The GBPMh results as presented in the main paper of all the clusters and liquid water, are based on the model parameterization using 10000 structures of (H$_{2}$O)$_{4}$ cluster along the trajectory of molecular dynamics at 50 K. However, to check the requirement of number of structures for the model parameterization of GBPMh, we have carried out a convergence test with the number of required structures of (H$_{2}$O)$_{4}$ at 50 K. We considered 40 and 100 structures selected randomly along the simulation trajectory of (H$_{2}$O)$_{4}$ at 50 K for the parameterization of GBPMh. We then used the parameters to calculate the GBPMh-polarizability for the whole trajectory of 30 ps with a time step of 0.003 ps (10000 structures). Fig.~\ref{Fig7} show the results of $\alpha_{xx}$ calculated using GBPMh and TDDFPT for the model parameterization using 40, 100 and 10000 structures. Fig.~\ref{Fig8} show the calculated Raman spectra for each of these cases. Clearly, the results using GBPMh-model parameterization based on 40 structures give good agreement with the TDDFPT results.           

\begin{figure}[h]
    \centering
    \includegraphics[width = 11.2 cm,angle =0]{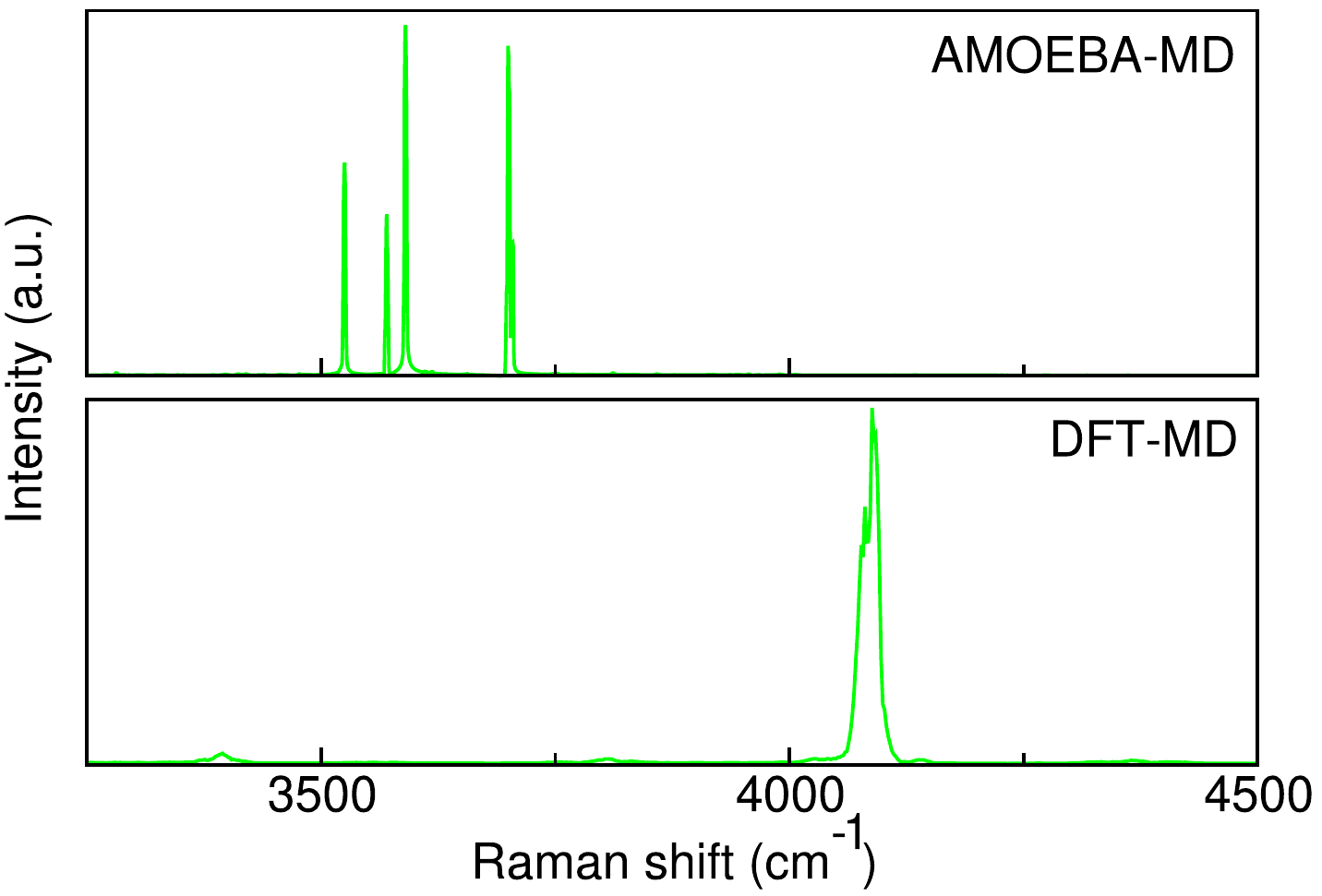}
    \caption{(H$_{2}$O)$_{4}$ at 50 K: Raman spectra using GBPMh, where MD is carried out using AMOEBA (top panel) and DFT (bottom panel).}
    \label{Fig9}
\end{figure}